\documentclass[12pt]{article}
\usepackage{graphicx,array, floatrow}
\usepackage{amsmath}
\usepackage[left=2.5cm, right=3cm, top=1cm]{geometry}

\usepackage{amsthm}
\usepackage{amssymb}
\usepackage{float}
\usepackage{array, booktabs}
\usepackage{indentfirst}
\usepackage{comment}
\usepackage{subcaption}
\usepackage{color,soul}
\usepackage{subfig}
\usepackage{algorithm}
\usepackage{algpseudocode}
\usepackage{amssymb,amsmath,amscd,amsthm}
\usepackage{graphicx,psfrag,epsfig}
\usepackage{graphicx,psfrag,epsfig, color ,float}

\usepackage{graphicx}
\usepackage[active]{srcltx}

\date{}

\newcommand{\RR}{\mathbb{R}}

\newcommand{\pt}{\partial}

\begin{document}

\date{}

\title{Inhomogeneous Branching Random Walks: Incorporating Genealogy and Density Effects}
\author{L. Ajax$^*$,
 B. Durham$^*$,
 P. Hebbar\footnote{Department of Mathematics, Grinnell College, Iowa, USA. {\it Correspondence:} hebbarpr@grinnell.edu},
   C. Johnson$^*$,
  J. Zhang$^*$}

\maketitle

\begin{abstract}
    
In this paper, we introduce a novel framework using inhomogeneous Branching Random Walks (BRWs) to model growth processes, specifically introducing genealogy-dependence in branching rates and displacement distributions to model phenomena like bacterial colony growth. Current stochastic models often either assume independent and identical behavior of individual agents or incorporate only spatiotemporal inhomogeneity, ignoring the effect of genealogy-based inhomogeneity on the long-time behavior of these processes. Such long-time asymptotics are of independent mathematical interest and are crucial in understanding the effect of patterns. We propose several inhomogeneous BRW models in 2D space where displacement distributions and branching rates vary with time, space, and genealogy. A combined model then uses a weighted average of positions given by these separate models to study the shape of the growth patterns. Using computer simulations, we tune parameters from these models, which are based on genealogical and spatiotemporal factors, observe the resulting structures, and compare them with images of real bacterial colonies.
\end{abstract}

{\it Key words: } Branching Random Walks, Inhomogeneous Branching Brownian Motion, Genealogy, Expanding Bacterial Colonies.

\section{Introduction}\label{introduction}

\subsection*{a. Motivation}

In an excellent review article of Jacob et. al. \cite{jacob}, we first came across several images of complex patterns that bacterial colonies can exhibit when faced with changing influencing factors, such as branching levels, nutrient availability, attractive and repulsive forces, and differing chemotactic strategies.

\begin{figure}[h]
\centering
\begin{subfigure}[c]{0.20\textwidth}
 \caption{}
\includegraphics[width=1.32in]{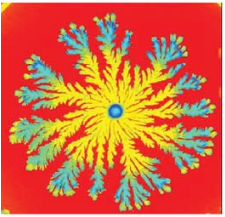}
  \label{one}
\end{subfigure}
  \begin{subfigure}[c]{0.20\textwidth}
 \caption{}
\includegraphics[width=1.32in]{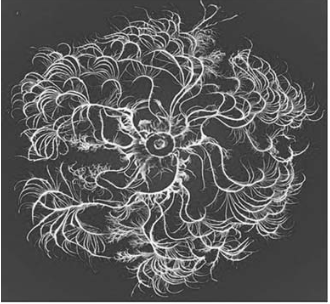}
  \label{two}
\end{subfigure}
  \begin{subfigure}[c]{0.20\textwidth}
 \caption{}
\includegraphics[width=1.32in]{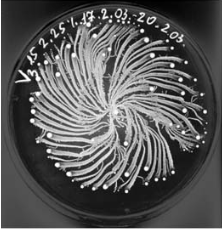}
  \label{three}
\end{subfigure}
  \begin{subfigure}[c]{0.20\textwidth}
 \caption{}
\includegraphics[width=1.32in]{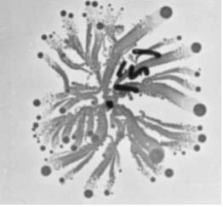}
  \label{four}
\end{subfigure}
\caption{All the images above are those of patterns of growing bacterial colonies subject to various external conditions. All these images were taken from \cite{jacob}.  In \cite{jacob}, Figure \ref{one} appears on page 242 Fig.2(a), Figure \ref{two} appears on page 243 Fig.3 (a), Figure \ref{three} appears on page 245 Fig. 5 (c), and Figure \ref{four} appears on page 246 Fig. 6 (e). We do not detail the species of each of the bacteria in each of the images since it is not relevant for our exposition below, but we refer an interested reader to \cite{jacob} for those biological details.}
\end{figure}

Computational simulations that focus on bacterial behavior from individual to population level have been studied \cite{biology11020297}, but authentic models that accurately reproduce real-life colony formations are rare and can be oversimplified \cite{MR2562651}. More commonly, partial differential equations have been used to describe the general characteristics and average behavior of bacteria, especially in large and homogeneous bacterial populations. These deterministic models require fewer computational resources and are able to capture the long-term behavior of the overall population more effectively. However, they are not necessarily the best fit when attempting to simulate bacterial behavior at individual levels. In contrast, a stochastic model can better reflect the behavior of individual particles as well as emergent population-level patterns, by allowing randomness and fluctuations that can significantly impact the entire system \cite{MR2562651}. Mathematical modeling and simulation are now a vital tool for understanding the complex, nonlinear dynamics of bacterial colony formation, working alongside traditional experimental methods. Simulations provide a crucial way to test hypotheses and explore scenarios that are difficult to replicate in a lab. These theoretical models, ranging from partial differential equations-based models to branching random walk models, have helped uncover the key drivers of colony expansion and pattern formation. These simulations can help predict the spread of pathogens and help study the influence of external and internal stimuli, including nutrient availability and intermolecular interaction forces on the shape, speed and pattern of the bacterial colony.

Random walk models have been widely used in biological literature, particularly in capturing the dynamics of bacterial colony growth. Individual bacteria are considered agents that move randomly through space and time while following certain rules for crowding, or interaction with each other, allowing more realistic simulations of cell migration and colony expansion \cite{SIMPSON20103779}. Similarly, branching processes where individual particles give birth to offspring according to a fixed (identical) offspring distribution play an important role in representing the stochastic reproduction of bacteria over time. As a result, Branching Random Walks (BRWs) are among the most commonly used models to simulate the growth patterns of bacterial populations.

Classical modeling in cell biology often considers BRWs in inhomogeneous media, where the inhomogeneity lies in the offspring distributions, and the displacement distributions due to variations in factors that are dependent on location (in space) and time. This is referred to as spatiotemporal inhomogeneity in BRWs.  One motivation for this study is the lack of genealogy-dependent factors in most existing mathematical models, which overlook the role that lineage-based patterns may play in the formation of bacterial colonies. We are interested in the role of each influencing factor separately, as well as how their integration influences the overall behavior of the model.

\subsection*{b. Overview of Branching Random Walks}

Imagine that there is a single bacterial cell sitting in a location $x_0\in \RR^2$; it is about to move through space. At each timestep, it moves a discrete step up, down, left, or right, with equal likelihood. Repeat this process, where each step is independent of the outcome of any other move. This elementary procedure is an example of a \textbf{simple random walk (SRW)} on a two-dimensional lattice, which is one of the most basic and widely studied stochastic processes \cite{Lawler2010}. Letting $S_n$ denote the position of the bacterial cell at time $n$, we can write
\begin{equation}
    \begin{aligned}
S_n=x_0+X_1+...+X_n       
    \end{aligned}
\end{equation}
where each $X_j$ is an independent random variable that describes the direction of the bacterial cell's $j$th movement, with $\mathbb{P}(X_j=\text{up})=\mathbb{P}(X_j=\text{any other direction})=\frac{1}{4}$.

Imagine again that this bacterial cell, while moving on the lattice, splits into a random number of new bacterial cells with a certain probability every timestep. Each of those offspring at the next timestep in the future splits again, independently. This is the idea behind the \textbf{Galton–Watson process}, a model that describes the growth of the bacteria population in discrete time, where each individual produces a random number of offspring \cite{Athreya_Ney}. The Galton–Watson process is one of the simplest examples of a \textbf{branching process}—a collection of models used to describe systems in which particles replicate or reproduce over time \cite{Athreya_Ney}. When the branching process is combined with a random walk, the result is a \textbf{branching random walk (BRW)}. In a branching random walk, at each timestep, every cell of bacteria takes a step and divides into $k$ offspring, with probability $p_k$, $0\leq k \leq \infty$, where $\sum^\infty_{i=0}p_i=1$.

Branching Random Walks with genealogy-dependent and spatiotemporal inhomogeneity that we consider have offspring distributions that depend on time, and displacement distributions (at time $t$) that depend on space, time, and on the underlying tree (from time $0$ to time $t$), where the genealogical structure itself is dictated by non-stationary reproduction laws. This structure allows for the formal study of how inherited traits, such as accumulated fitness through successive mutations, affect the subsequent growth and dispersal dynamics.

\begin{figure}[h]
    \centering
    \includegraphics[width=0.35\linewidth]{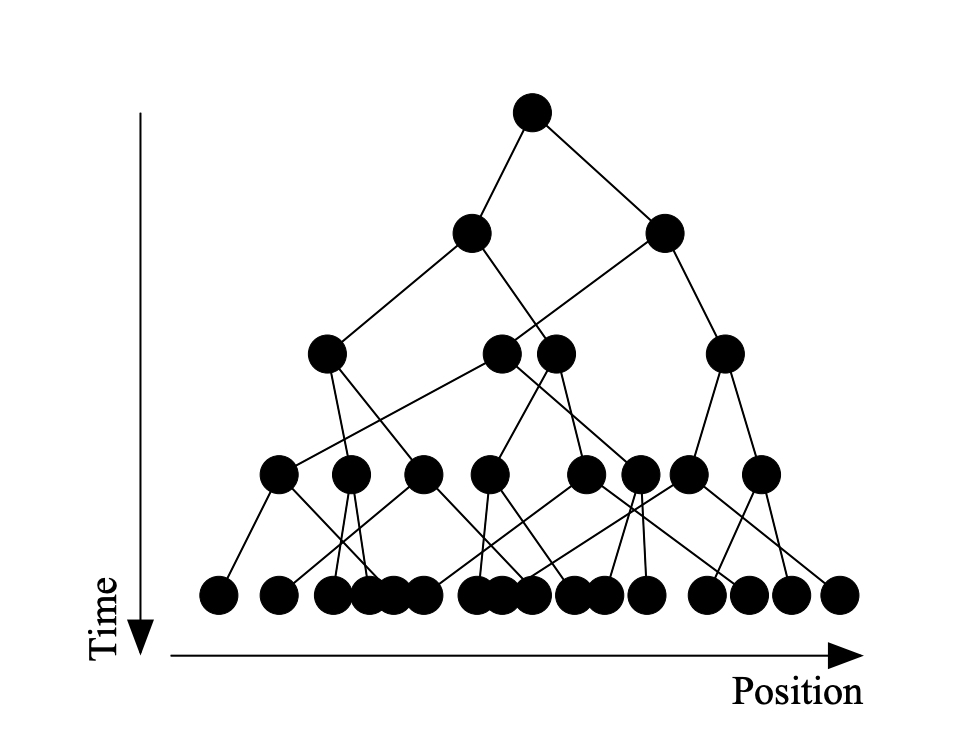}
    \caption{A branching walk in discrete time where each particle splits into two children at each time step: $p_2$=1. \cite{MR2799946}}
    \label{fig:enter-label}
\end{figure}

Let us briefly discuss the connection between the discrete-time simple random walks and branching random walks, to the continuous-time Brownian motions and branching Brownian motions. Let $d = 1$ for simplicity. It is well understood that as the lattice spacing $\delta$ and timestep $\tau$ become infinitely small or tend to zero, if the ratio of space and time satisfies 
\begin{equation}
    \begin{aligned}
    \frac{\delta^2}{\tau}\rightarrow 2D, \text{\;where $D>0$ is some constant, later defined to be the diffusion coefficient,}  \end{aligned}
\end{equation}
then discrete random walk converges to Brownian motion \cite{donsker1951invariance}, which is considered the continuous-time counterpart of the random walk \cite{donsker1951invariance}. As $\delta, \tau \to 0$, probability density of the location of the particle $u(x, t)$ undergoing a SRW starting from location $x_0$, is the solution to the heat equation, also known as the Fickian diffusion equation \cite{lin1974mathematics}.

\begin{equation}
    \begin{aligned}
\frac{\pt u}{\pt t}=&\underbrace {D}_{\text{ Diffusivity}} \cdot \; \frac{\pt^2 u}{\pt x^2}\\
u(0,x) = & ~\delta(x-x_0)\\
    \end{aligned}
\end{equation}
\begin{center}
    Equation 2: {\footnotesize The heat equation describes how heat diffuses over time in one dimension.}
\end{center}

While the heat equation successfully describes the diffusive behavior of non-interacting and unbiased particles undergoing Brownian motions, it does not account for systems that involve the reproduction or annihilation of particles; here, the density of particles is modeled by the following heat equation with potential 
\begin{equation}
    \begin{aligned}
\frac{\pt u}{\pt t}=&\underbrace {D}_{\text{ Diffusivity}} \cdot \; \frac{\pt^2 u}{\pt x^2} + \underbrace {v}_{\text{Branching}} u  \\
u(0,x) = & ~\delta(x-x_0)\\
    \end{aligned}
\end{equation}
where the function $v$ represents the branching rate of the particles. This process is also related to the renowned reaction-diffusion equations. Observed by McKean in 1975, the solution $u(t,x)$ of a specific reaction-diffusion equation named Fisher – Kolmogorov – Petrovsky – Piskunov (FKPP) can be interpreted as the probability that no particles have reached position $x$ by time $t$ in a system of branching particles undergoing Brownian motions \cite{Mckean1975}. Refining and extending McKean's probabilistic approach, Bramson proved that the rightmost particle in the long-time limit can be described by the solution of the FKPP equation \cite{bramson1983convergence}, \cite{https://doi.org/10.1111/j.1469-1809.1937.tb02153.x}, \cite{MR2838339}. Let $n(t)$ denote the number of particles alive at time $t$ and let $x_k(t)$ denote their locations for  $1\le k\le n(t)$.

\begin{equation}
    \begin{aligned}
\frac{\pt u}{\pt t}&=\underbrace {D\frac{\pt^2 u}{\pt x^2}}_{_{\text{ diffusion term}}} + \underbrace {f(u)}_{\text{reaction term}}\\
\end{aligned}
\end{equation}
\begin{equation}
\begin{aligned}
\text{where}\; 
0\le u(t,x)&=:\mathbb{P} [\text{max}_{1\le k\le n(t)} x_k(t) \le x]\le 1 
    \end{aligned}
\end{equation}
where $u(t,x)$ represents the probability that the maximum of the positions of the particles undergoing branching Brownian motion is less than $x$ at time $t$ (see \cite{MR3615508} for reference).

 The FKPP is a reaction-diffusion equation that describes the diffusion of particles over space and time. If the branching rate is a constant $v$, the corresponding reaction or the growth term in FKPP is typically $f(u)=v u(1-u)$, where $u$ represents the population density. Recall that the heat equation is defined by $\frac{\pt u}{\pt t}= D \frac{\pt^2 u}{\pt x^2}$, which is a special case of the FKPP equation where $f(u)=0$. The FKPP equation has two homogeneous steady-state solutions: while the unstable solution $u=0$ represents the extinct state, the stable solution $u=1$ corresponds to the saturated or fully populated state, where the population has reached its maximum density (see \cite{MR4414125} \cite{MR3615508} for further details).

The equations introduced above provide a solid mathematical backbone for this paper: they can describe how a colony's boundary moves outward over time as the population spreads, forming a traveling wave. Moreover,
most complex behavior of bacteria (including time-inhomogeneous branching functions, genealogical repulsion or adhesion, etc.) can often be simply viewed as perturbations to the PDEs, caused by particle interactions, variability in parameters, or environmental factors. This coordinated communication allows the colonies to collectively make decisions and alter their structure, suggesting a form of "social intelligence" that may also involve passing down a shared memory.

\subsection*{c. Previous Research}

While branching Brownian motion in homogeneous media has been studied extensively, generally through the lens of partial differential equations \cite{pastbbmexample1}, \cite{Plank_motiv_paper}, \cite{pastbbmexample2}, there has been less research on branching Brownian motion in inhomogeneous media. There has been research into branching Brownian motion with an inhomogeneous branching rate based on position \cite{inhombbm1} or otherwise \cite{inhombbm2}, and into time-inhomogeneous branching Brownian motion \cite{inhombbm3}, but there has not yet been much, if any on the subject of branching Brownian motion that is inhomogeneous due to genealogy. This is in part due to the immense difficulty of mathematical analysis of this subject, as we are allowing the rules of the branching Brownian motion model to be changed by the genealogy of each particle. We therefore attempt to analyze this problem through computer modeling simulations, playing with genealogical and spatiotemporal factors introduced in Section \ref{influencing factors} and observing the resultant structures, comparing and contrasting them with images of bacterial colonies.  

When modeling, the population size and therefore number of data points grows exponentially, in our case on the order of $2^t$, and as such, the technology accessible to us limited our analysis to the first ten to twenty generations. There has not been much coding research dedicated to this question, and as we were unable to find a workaround, our conclusions apply only to the initial few generations, and we are unable to discuss the effects of the influencing factors on the limiting shape of the random walks.

The paper is organized as follows: in Section \ref{influencing factors}, we introduce the major influencing factors affecting the formation of bacterial growth and how we utilize each of those factors in our model. In Section \ref{model descriptions}, we describe all the models we propose and simulate, and discuss how different functions of the radius of influence affect the genealogy-based cloud and cluster model in Section \ref{function types}. In Section \ref{findings}, we display our findings, while in Sections \ref{conlcusions} and \ref{further research}, we discuss our conclusions and directions for future research.

\section{The Influencing Factors}\label{influencing factors}

We consider two main categories of influencing factors, genealogical and spatiotemporal, introduced below. All of the models discussed in Section \ref{model descriptions} implement a combination of these two categories.

\subsection*{a. Genealogical Factors}
Genealogical relationships have been shown to exert a measurable influence on the spatial structure of bacterial populations. Biological studies have demonstrated that interactions between clonal lineages can directly impact overall structure and the growth pattern of the colony \cite{Kan2018}. For example, experimental evidence indicates that closely related individuals, such as siblings, often physically adhere, while more distantly related particles tend to become spatially separated over time \cite{Kan2018}. 

Genealogical factors were incorporated into branching random walk models mimicking the behavior of bacterial colony growth patterns by controlling how the model was allowed to grow. By making sure the identity of the parent was stored in the code and easily accessible by its children, new nodes are able to use the information of all their ancestors to determine their behavior. A child could be limited to traveling only within a certain distance from its parent, or be required to stay within a certain angle of its siblings. In more elaborate cases, the number of children its previous five ancestors had could be averaged to determine its own number of children, or, depending on whether its parent had more second cousins to its left or its right, it would determine its leaning. We found that the most practical use of genealogy was in letting the angle at which the parent node was determine the range in which all of its children's angles would fall. By restricting this range, the branching of the model would almost always cause children to move in an outward direction, enabling branches to more easily form and preventing the model from falling in on itself. Including genealogical factors enables models to grow into something that can be built on to create patterns akin to bacterial growth.

\subsection*{b. Spatiotemporal Factors}

Space-affected models have been studied before, such as in Plank et. al. 2025 \cite{Plank_motiv_paper}, where space exclusion models were discussed. In their models, agents are not allowed to inhabit the same space as any other agent. Our models use space differently from theirs. In our models, space is considered in the density aspect. In our models, higher densities lead to more outward angles of propagation, while less dense regions have fewer restrictions on the angle and direction of the children. 

For bacterial growth, the birthrate is very high in the initial few generations, causing the population to grow very quickly. To mimic this, we discussed allowing the first few generations' birthrates to be very high, up to 10 children per node. We decided against this model because the growth rate being on the order of $10^n$ would greatly increase the runtime of the program while simultaneously reducing the number of generations we could feasibly produce. Throughout, we were limited by the capabilities and memory space of our technology. 

Another possible effect of density on the model is through birthrate. We discussed a model wherein the regions of high density correspond to regions of high birthrate, while regions of low density likewise correspond to regions of low birthrate.

\section{Model Descriptions}\label{model descriptions}

\begin{figure}[]
\centering
    \caption{Simulation of various models}
    \begin{subfigure}[c]{0.30\textwidth}
  \caption{Simple Random Walk on a 2D lattice.}
    \includegraphics[width=\textwidth]{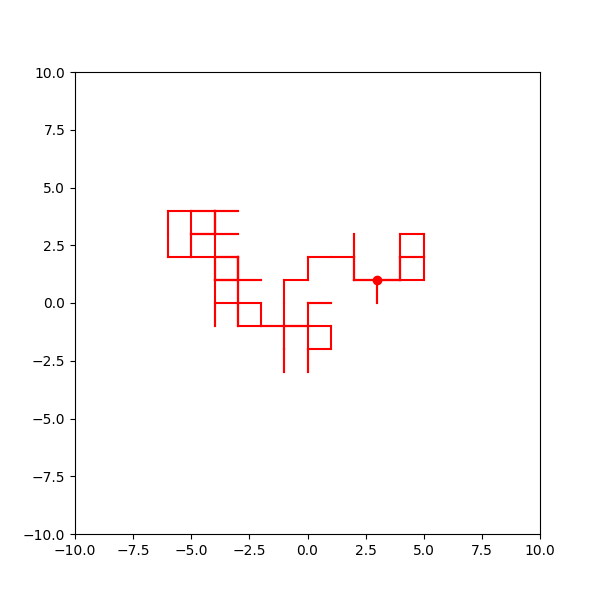}
    \label{fig:enter-label}
    \end{subfigure}\hfill
    \begin{subfigure}[c]{0.30\textwidth}
     \caption{Branching Random Walk on a 2D lattice}
        \includegraphics[width=\textwidth]{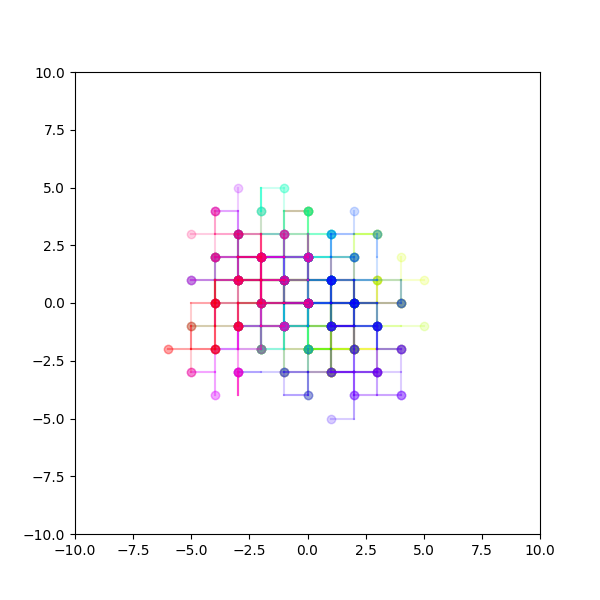}
  \label{fig:BRWLat}
    \end{subfigure}
    \begin{subfigure}[c]{0.30\textwidth}
      \caption{Heat diffusion visualization\\
      \
      \\
      \
      }
         \includegraphics[width=\textwidth]{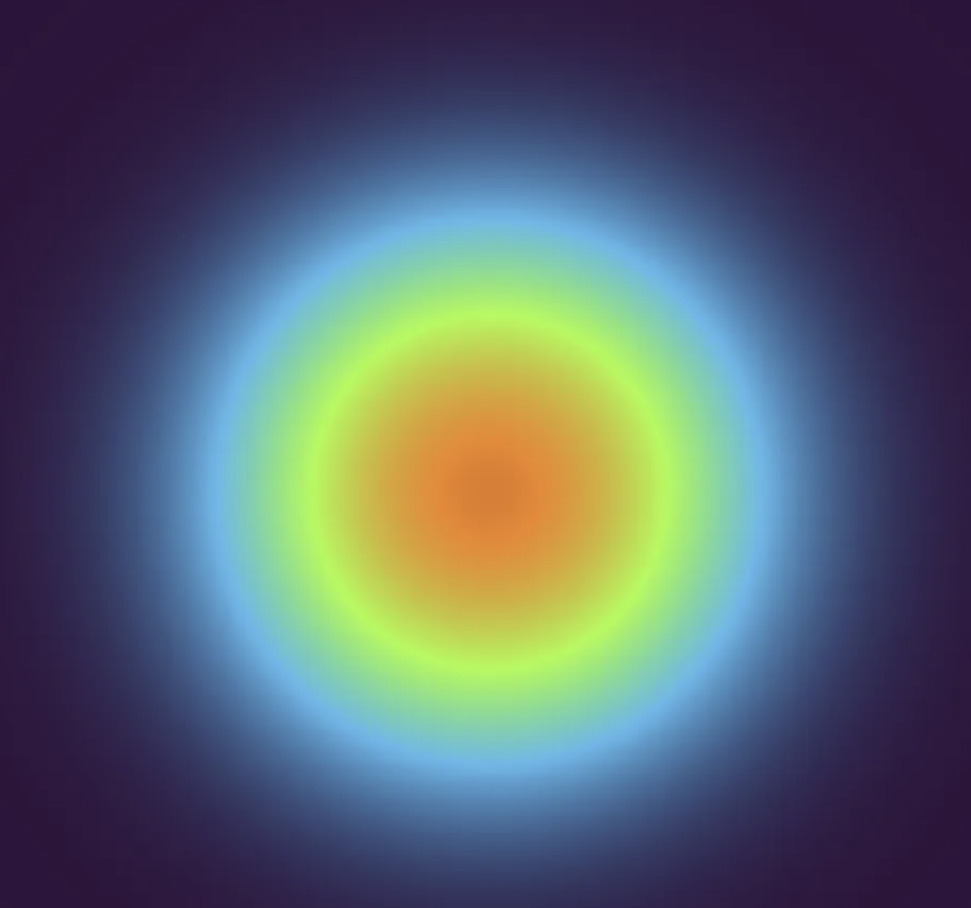}
  \label{fig:heatdiff}
    \end{subfigure}
    \begin{subfigure}[c]{0.30\textwidth}
  \caption{Random Model with each node's direction being completely arbitrary.}
    \includegraphics[width=\textwidth]{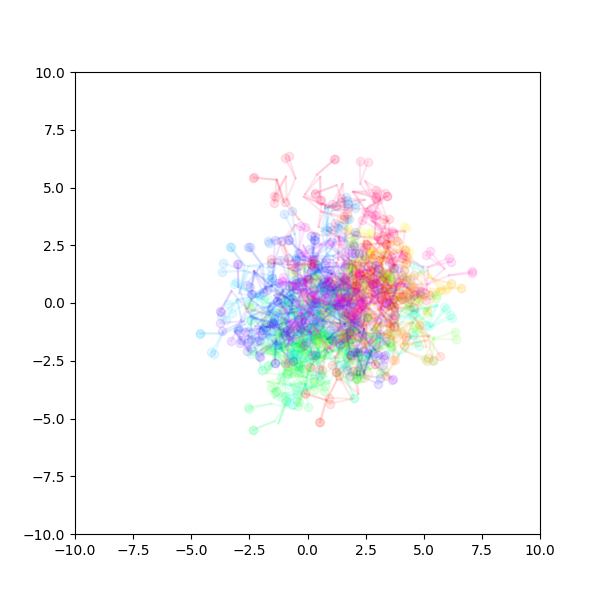}
    \label{fig:enter-label}
    \end{subfigure}\hfill
   \begin{subfigure}[c]{0.30\textwidth}
        \caption{ Bias Model with significantly increased odds for its nodes to travel east, northwest, or southwest.}
    \includegraphics[width=\textwidth]{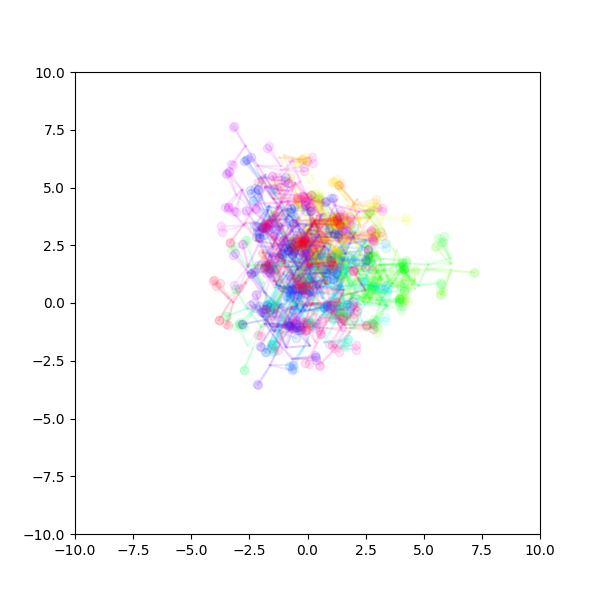} 
    \end{subfigure}\hfill
    \begin{subfigure}[c]{0.30\textwidth}
       \caption{Parent Influenced Bias Model whose parent nodes influence their children's positions.}
    \includegraphics[width=\textwidth]{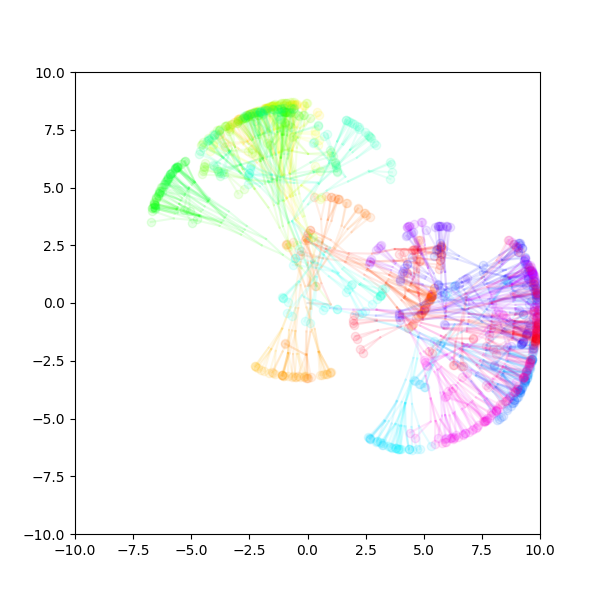}
    \label{fig:tranglebranchingparentangle}
    \end{subfigure}\hfill
    \begin{subfigure}[c]{0.30\textwidth}
     \caption{Cloud Model that repels nodes when they get within a set distance from each other.}
    \includegraphics[width=\textwidth]{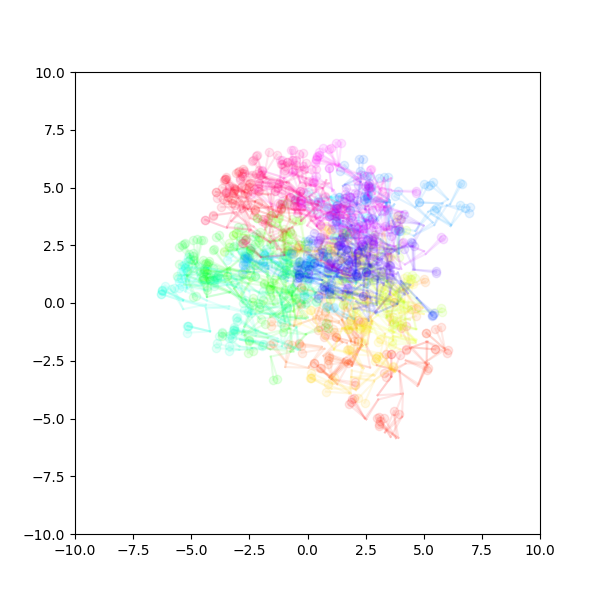}
    \label{fig:enter-label}
    \end{subfigure}\hfill
     \begin{subfigure}[c]{0.30\textwidth}
    \caption{Density model with nodes in largely dense areas traveling in the same direction.}
 \includegraphics[width=\textwidth]{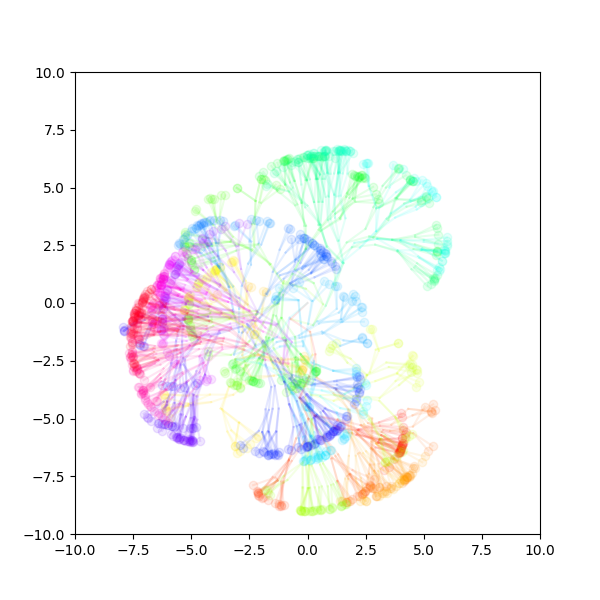}
    \label{fig:enter-label}
    \end{subfigure}\hfill
   \begin{subfigure}[c]{0.30\textwidth}
     \caption{Cluster model that clusters nodes and moves them in the same direction.}
        \includegraphics[width=\textwidth]{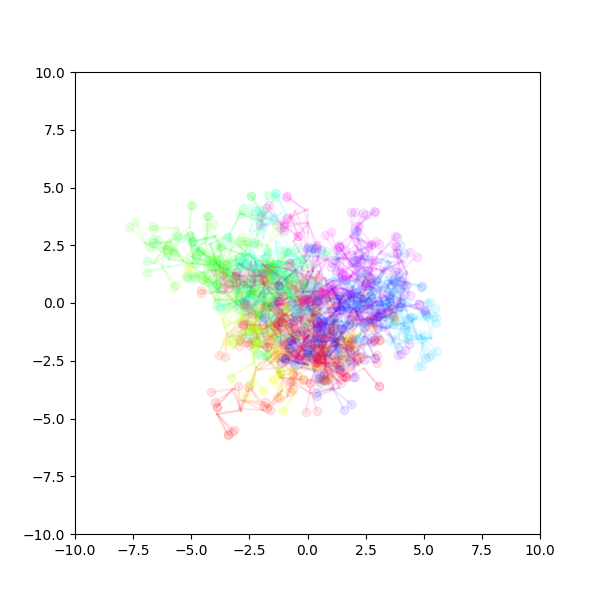}
    \end{subfigure}\hfill
    \label{img1&2}
\end{figure}

This section covers the different models that we attempted in our effort to model bacterial growth. Several models were constructed using a branching random walk framework with various influencing factors included, as discussed in the previous section. To begin, we constructed individual models and ultimately built a combined model that enabled all the mentioned models to be either run individually or be in weighted combinations. The combined model would take variety of parameters including the weight of each model, the angle at which the parent influenced bias model's children would be allowed to vary from their parent's angle, the angle at which siblings would be allowed to vary from each other, the angle at which children would be allowed to vary from the cluster model's guiding angle, the angle at which the bias's models children would be allowed to vary from their selected direction, the number of children each node would have, the number of generations the model would run for, and how many of those generation would be random. The major flaw of this model was its computational complexity. Since all nodes needed to be stored in order to track genealogy, and the number of nodes would increase exponentially if the number of children born to each node exceeded one, a necessary step in observing growth, the program would quickly run out of space. Thus, we had to limit, for the most part, the number of generations we ran to ten, and therefore, we could not observe long-term behavior. The following section will dive into the models as we originally attempted them, how we constructed them in the combined model, and what our goal was for each model. Further experimentation with the combined model can be seen in Section \ref{findings}. 

\subsubsection*{Model Coloring}
Before the models, we would like to write a note about the coloring of the models. The first node created is assigned a color, and when it has children, one of its children retains the same color, while all the others receive a color that is slightly different. The result is that the more similar the color of two nodes is, the more closely they are related.

\subsection*{Simple Random Walk on a Lattice}

\textbf{Simple Random Walk (SRW) Models} are elementary models that have their nodes move along a square lattice. The node will take a step, produce a child, select the direction the child will take, and then the child repeats the process. The four directions, up, down, left, and right, have an equal probability of being selected. For the sake of simplicity, each node only had one child (no branching), resulting in this model being able to sustain many more generations than models with branching, as the population size stayed constant at one and did not grow exponentially. The goal for this model was simply to familiarize ourselves with the concept of random walks, as well as to work out how such a thing would be implemented in code.

\subsection*{Branching Random Walk Model on a Lattice}

\textbf{Branching Random Walk Models} are classical stochastic processes that allow for the growth of the population undergoing SRWs. Since we are aiming to test multiple models in the effort to best model bacterial growth, we needed to figure out how to model the population growth. Therefore, we first study the classical approach, which is to take the SRW model and allow for two children to be born to each node at each time step. These children would independently select one of the four directions of the lattice to travel with equal probability. This model demonstrates how rapidly the population would expand, and limit our generation runtime once it does, and provides a solid backbone for branching models. An interesting note is that the limiting behavior of branching random walks can be modeled by the diffusion equation with a linear potential, as can be seen in the similarities between the two figures given in Figure \ref{fig:BRWLat} and \ref{fig:heatdiff}.

\subsection*{Branching Random Walk Model}

\textbf{Branching Random Walk Models} were constructed to adjust to moving off the lattice. The idea was that, rather than simply picking one of four directions, now our model's nodes could travel in 360 degrees, completely arbitrarily. However, there was one restriction. All sibling nodes would have to be within a distance set by the user of each other. We hoped that by observing completely random travel, we could get a better idea of what we had to do to best model bacterial paths. Additionally, this model could be used in conjunction with future models to allow for some variation in node position during the first few generations, to prevent the restrictions imposed by later models from being overly restrictive.

\subsection*{Bias Model}

\textbf{Bias Models} selects arbitrary biasing angles, for example, 0, 120, and 240, and chooses a parameter that determines the magnitude of the difference in angle, measured in radians, between the displacement vectors of two sibling nodes, with both vectors starting at their shared parent. Additionally, a small chance is added for a node to move in a completely random direction instead. By doing this, we were able to learn how to form the model into a very distinct shape and acquire some idea for how we could evolve it into something more natural. This model also incorporates the first few generations where nodes homogeneously undergo random walks without biasing.

\subsection*{Parent Influenced Bias Model}

\textbf{Parent-influenced bias models} were created to replicate the phenomenon of bacteria moving away from their origin point, as seen in bacterial growth. While the first generation of child nodes were assigned their directions based on the bias model, each subsequent generation of nodes would move in the same direction as their parents, with some variation according to user input. This enabled us to gain a solid understanding of how parent nodes influence their children, which helps us improve a model's expansion, similar to observed bacterial growth. All of the nodes are inclined towards moving outward in the same direction as the node that preceded them. They cluster around the boundary of the model in a ginkgo leaf shape.

\begin{table}[]
\begin{center}
\begin{tabular}{ | m{2.5cm} | m{9cm}| m{3.5cm} | } 
    \hline 
     \textbf{Parameters}  & \textbf{Description} & \textbf{Used in following models:}  \\
     \hline
     
       Generations  & The number of generations (time steps)  the model will run for.&  ALL\\
           \hline
       Children & The number of new nodes each (parent) node creates in the next generation & ALL but Simple Random Walk\\
           \hline
       Random Generations & The number of generations that will run completely randomly before the respective model algorithm is applied. & Bias, Parent Influenced Bias, Cloud, Density, Cluster, Combined.\\
           \hline
           Sibling Variance&  Magnitude of the difference in angle, measured in radians, between the displacement vectors of two sibling nodes, with both vectors starting at their shared parent.& Branching Random Walks, Bias, Parent Influenced Bias, Cloud, Density, Cluster, Combined.\\
            \hline
            Bias Variance &  Maximum allowed angular deviation (measured in degrees) of a node's displacement vector from a specific, predetermined bias angle. & Bias, Parent Influenced Bias, Combined.\\
            \hline
            Parent Variance & Magnitude of the difference in angle, measured in radians, between the displacement vectors of the child node (starting from the parent node) and that of the parent node (starting from its parent node).& Parent Influenced Bias, Combined.\\
            \hline
Radius of cloud of repulsion function $f(t,s)$ &  The radius of the ball around nodes in generation $t$ such that all the other nodes with most recent common ancestors with that node that are $s$ time in the past (referred to as $(s-1)-$th cousins) are forced to move away.  & Cloud, Combined. \\
            \hline
   Density Reinforcement & The weight that is placed on the local density while determining the displacement vector of a node. & Density, Combined.\\
   \hline
   Cluster size $f(t)$ & The radius of the ball around nodes in generation $t$ such that all the other nodes within that ball would be assigned the same guiding angle. & Cluster, Combined.\\
   \hline
   
   Cluster Variance &  Magnitude of the difference in angle, measured in degrees, between the displacement vectors of nodes, starting at their respective parent that are in the same cluster.& Cluster, Combined.\\
   \hline

    \end{tabular}
    \caption{Parameters used in the models.}
    \label{tab:placeholder}
    \end{center} 
\end{table}

\subsection*{Cloud Model}

\textbf{Cloud models} prevent nodes from accumulating together, based on the idea of spatial exclusion. Mimicking the branching patterns seen in bacterial growth, this model works by repelling nodes from one another based on their proximity and genealogical relationship. Initially, we set a constant repulsion spatial radius of 0.1. If any node entered this distance from another node, it would be forced to move away. We also modified this model to allow for a dynamic radius of repulsion. These functions took the degree of "cousin" (most recent common ancestor) relation between two nodes or the current generation number as input and returned a new radius. Some of these functions were designed to increase the radius as the relation became closer (as the most recent common ancestor became closer in time), while others decreased it. Section \ref{function types} provides a more in-depth discussion of these repulsion functions. As cousins are pushed away from each other, the model forms a star-like shape. The colors of nodes, and therefore families of nodes, stick close to each other.

\subsection*{Density Model}

\textbf{Density Models} replicate the way our bacteria samples formed distinctive, dense branches with barren gaps in between. This model works by first calculating the local density of an area. It does this by counting the total number of nodes within a radius of one (a constant we used throughout our experiments) and then dividing that total by four. This value is used to influence the position of a new node. If a node is in a high-density area, it is encouraged to move further in that direction. Conversely, if it is in a low-density area, the model moves it away, presumably toward a more concentrated region. This refined calculation method proved far more effective than our initial approach, which simply counted the number of nodes within the radius. The nodes in dense areas are traveling outwards faster. Meanwhile, nodes in higher-density areas fan out less while nodes in lower-density areas fan out more, once again reminiscent of a ginkgo leaf shape.

\subsection*{Cluster Model}

\textbf{Cluster models} aim to group nodes, to form distinct branches, similar to bacterial growth patterns. A function $f(t)$ generates a radius of clustering based on generation. All nodes within this radius of each other would be assigned the same guiding angle. Then, all those nodes' children would be within a variation from the guiding angle (as measured from their individual parent). We start with $f(t) = 0.012\times(\frac{2t+1}{t+1})$, as our radius function, because it was an equation that increased at a decreasing rate, with its limit preventing it from getting out of control, while the constant $0.012$ kept the overall value small enough. Section \ref{function types} contains further details about this process. While this model diffuses into a limited shape, from the colors we observe that, inside this spread, genealogically close families have stayed closer together than before. It fills the space homogeneously yet maintains closeness between genealogically close family members.

\subsection*{Combined Model}

The \textbf{Combined Model} integrated all previously developed individual models by computing the positions of the nodes predicted by each previous individual model and computing a new predicted position by taking a weighted average of those positions. The goal of this model is to observe the interplay and mutual effects of these components, allowing us to refine our overall approach. We systematically analyzed the simulation results by varying the weights assigned to each constituent model. These are explored in more detail in section \ref{findings}.

\section{Function Types}\label{function types}

In the previous section, the cluster and the cloud models' radii of attraction/repulsion were determined by functions. In the effort to find the best possible functions for our models, we experimented with multiple equations that increased, decreased, or stayed constant over time. This section will discuss the functions that were implemented in the models.

\begin{figure}[H]
\caption{Various Functions considered in Cloud and Cluster Models for dynamic radius computation.}
 \begin{minipage}[m]{.50\textwidth}
\centering
\begin{tabular}{|l|c|r|}
\hline
\textbf{Function} & \textbf{Definition} & \textbf{Range} \\
\hline
Exponential Decay         & $(9/10)^t$           & $(0,1]$             \\
Exponential Decay           & $e^{-t/5}$           & $(0,1]$             \\

Logarithmic Function& $\ln(t)/2$ & $(-\infty,\infty)$\\
Square Root Function& $\sqrt t/2$ & $[0,\infty)$\\
Logarithmic Function &$\frac{2t+1}{t+1}$ & [1,2)\\
Linear Function       & $t/10$           & $(-\infty,\infty)$              \\
Constant Function & 1/10 & $\{1/10\}$\\

\hline
\end{tabular}
\label{tab:sample}
\end{minipage}
\begin{minipage}[m]{0.45\textwidth}
      \centering 
    \includegraphics[width=2in]{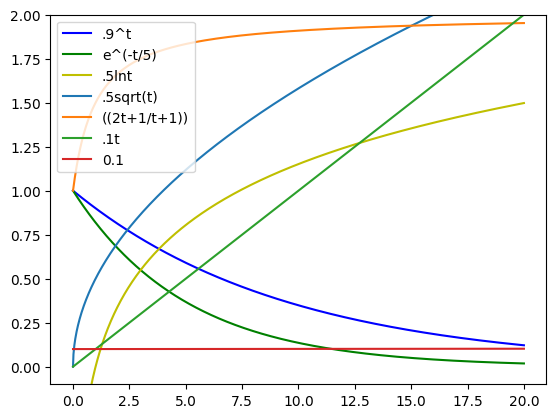}
    
    \label{fig:enter-label}
\end{minipage}
\end{figure}

\subsection*{Cluster Model Functions}
The Cluster model was designed to aggregate the progeny of spatially proximal nodes. Establishing a formal definition for "proximal" necessitated the use of the generation number $t$ as a key parameter to define the clustering radius. Our initial hypothesis posited that increased branching over time would necessitate a progressively smaller radius to maintain meaningful cluster formation. We evaluated several time-dependent, decreasing functions (scaled to fit our model) to define this radius, including $e^{-t}$, $(9/10)^t$, and $(-\ln(t+2)+3)/2$. We tracked the number of clusters per generation to monitor the model's behavior. A critical flaw was swiftly identified: as $t\to \infty$, these functions converge to zero. This convergence caused each node to eventually form its own isolated cluster, thereby negating the intended purpose of the aggregation mechanism and preventing the desired clustering behavior. Consequently, we shifted our focus to time-dependent, increasing functions (scaled to fit our model), such as: $t/10, \ln(t)/2$, and $\sqrt{t}/2$.  However, this approach introduced a different instability. Without a defined upper boundary, the clustering radius expanded without a bound. This expansion led to a systematic reduction in the number of clusters, culminating in the coalescence of the entire model into a single cluster.

\begin{figure}[H]
\caption{Images of cluster model run with various equations for the radius of attraction, along with the cluster count by generation.}
\begin{subfigure}{0.49\textwidth}
\includegraphics[width=0.9\linewidth, height=4cm]{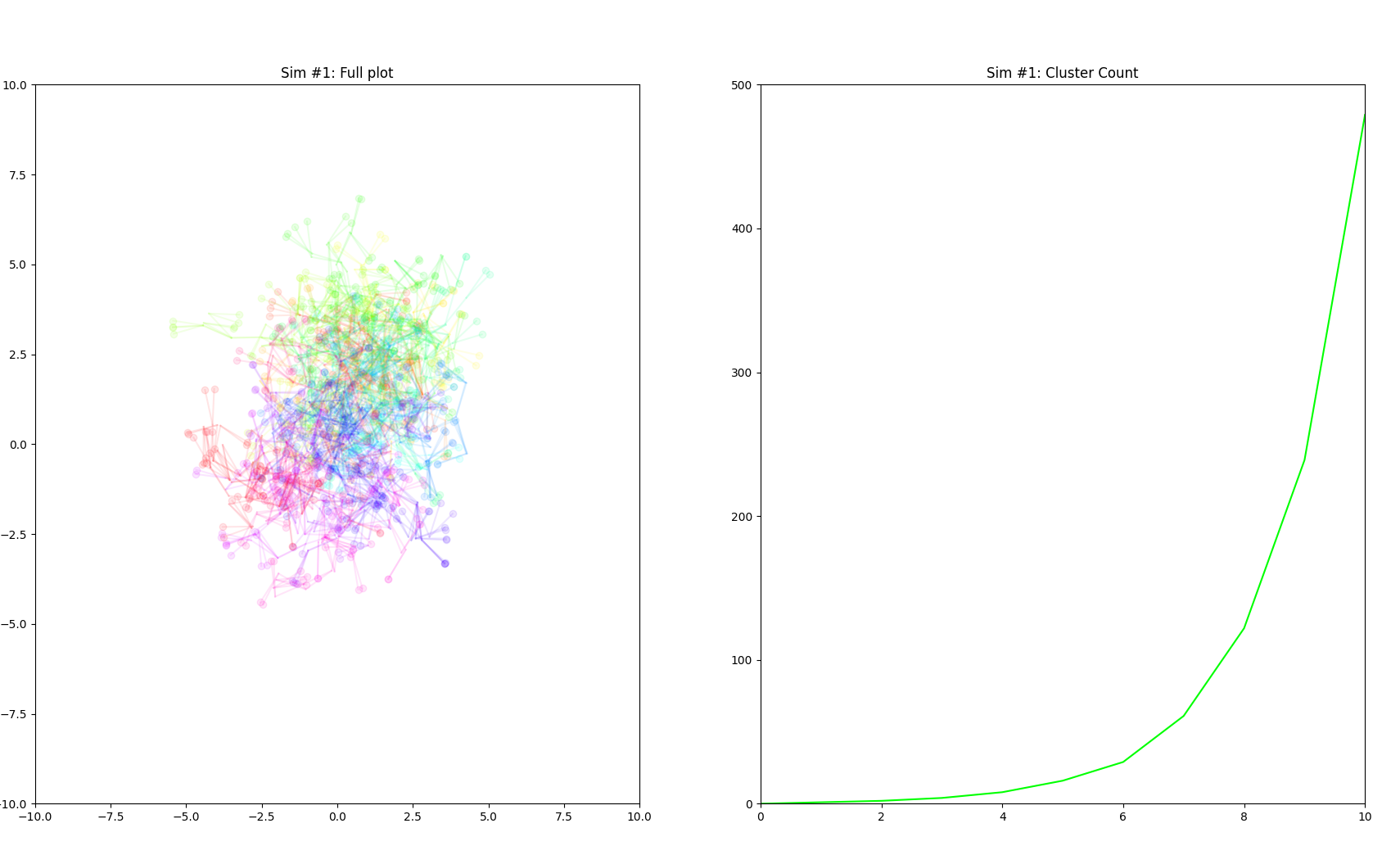} 
\caption{ Radius function $f(t) = 0.012(\frac{2t+1}{t+1})$}
\label{fig:subim1}
\end{subfigure}
\begin{subfigure}{0.49\textwidth}
\includegraphics[width=0.9\linewidth, height=4cm]{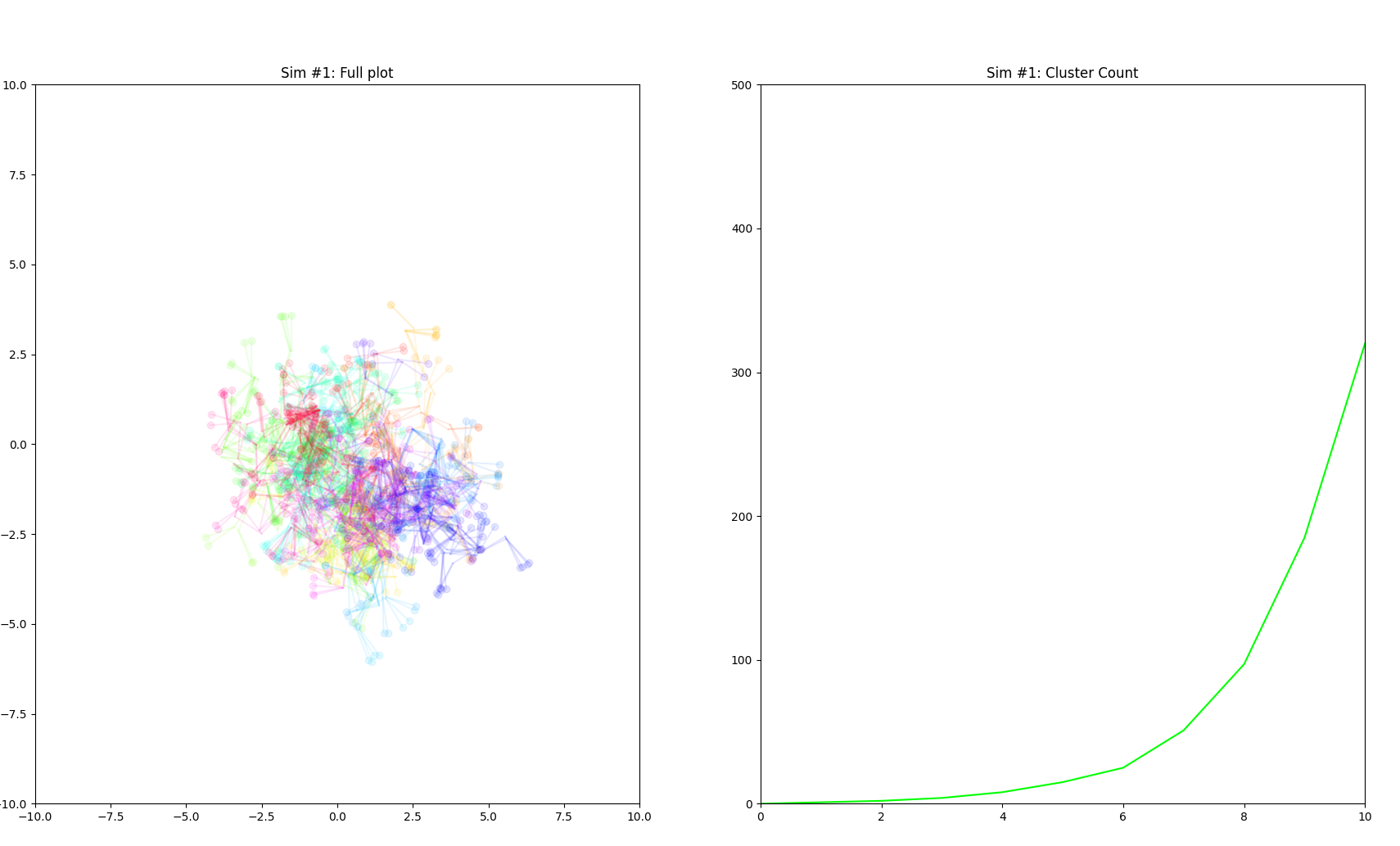}
\caption{ Radius function $f(t) = t/10$}
\label{fig:subim2}
\end{subfigure}
\begin{subfigure}{0.49\textwidth}
\includegraphics[width=0.9\linewidth, height=4cm]{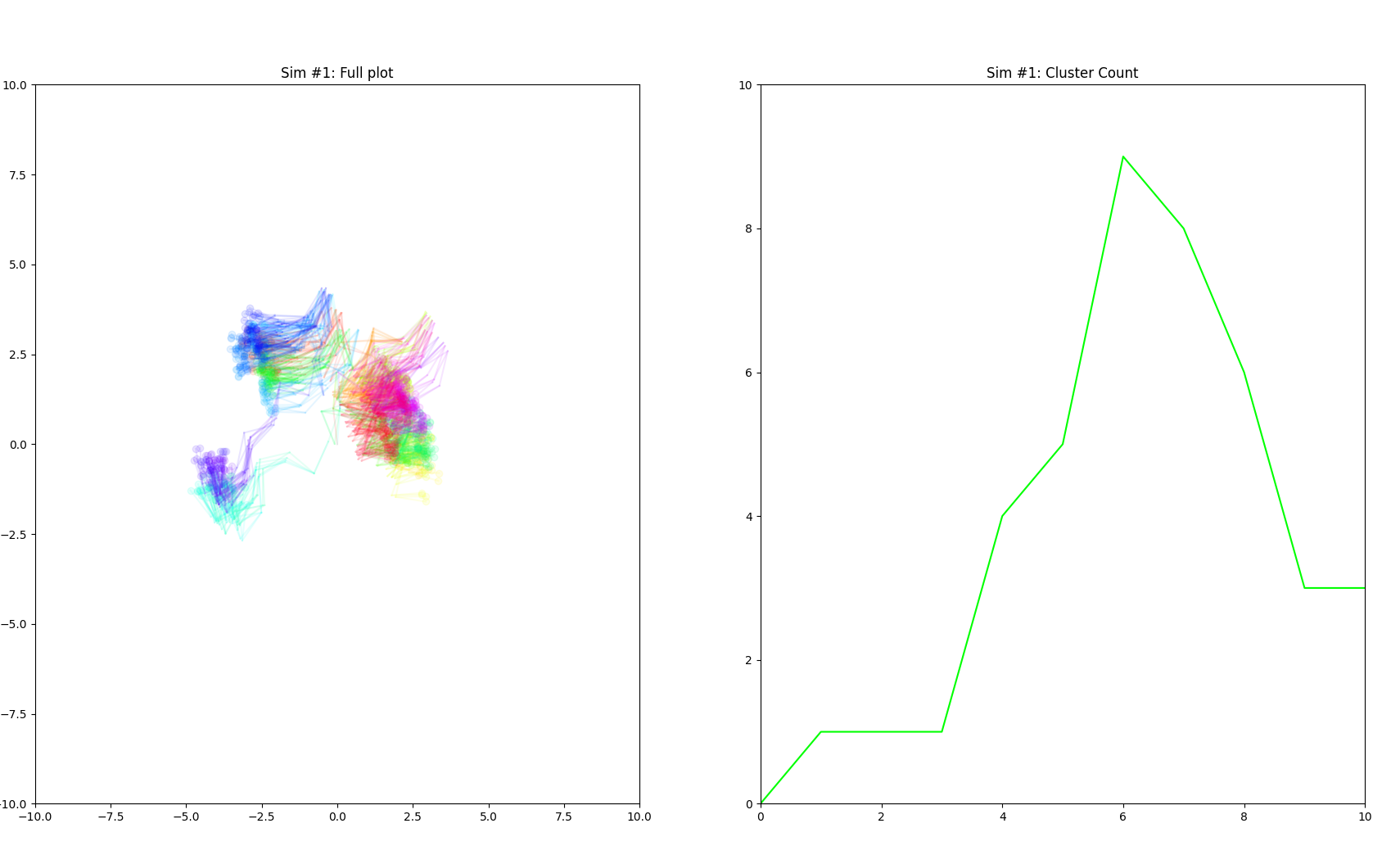}
\caption{ Radius function $f(t) = (9/10)^t$}
\label{fig:subim3}
\end{subfigure}
\begin{subfigure}{0.49\textwidth}
\includegraphics[width=0.9\linewidth, height=4cm]{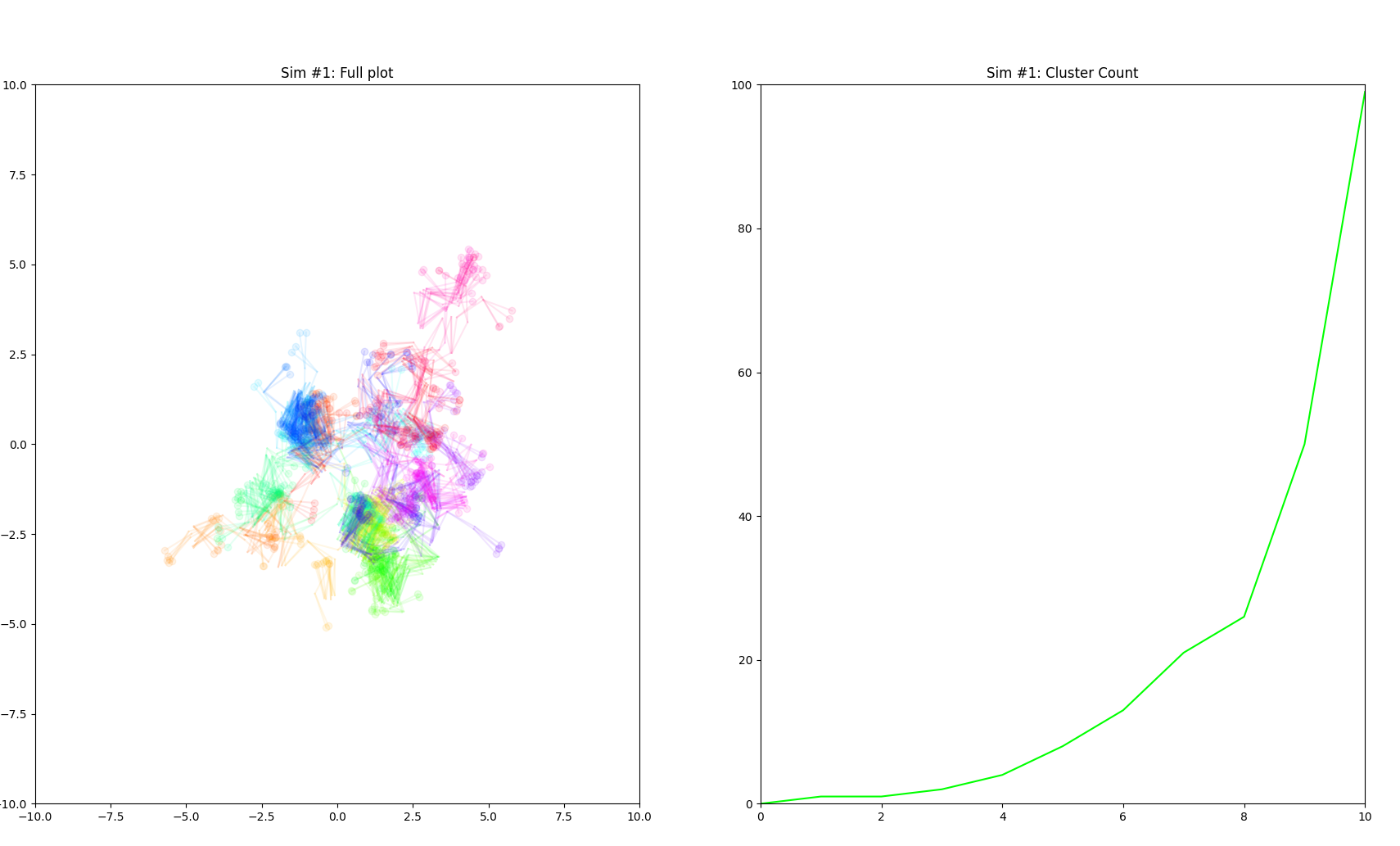}
\caption{ Radius function $f(t) = e^{-t/5}$}
\label{fig:subim4}
\end{subfigure}
\label{fig:image1}
\end{figure}

What we needed was a function that increased at a decreasing rate and had a limit. This led us to our final model, which uses the base equation $\frac{2t+1}{t+1}$. This function provided a rate of increase that worked well with our model and had a limit that was not too far from its starting point, allowing for both clustering and branching.

\subsection*{Cloud Model Functions}

The objective of the Cloud model was to implement a mechanism for repulsion between nodes to discourage the proximity of particles. This repulsion mechanism was defined by a radius, the size of which was the subject of significant experimentation. The initial hypothesis suggested that repulsion should be maximized early in the simulation to allow initial nodes to establish distinct trajectories before subsequent generations could cluster. This was tested using time-dependent, decreasing functions for the repulsion radius, such as: $9t/10$ and $e^{-t/2}$.
However, this strategy proved detrimental to the model's structure. The strong initial repulsion created large, persistent spatial voids that later generations were unable to fill, resulting in model formations that were artificially sparse and incomplete. We then shifted our focus, assuming that the repulsion radius should increase over time. Furthermore, we considered making the radius dependent on the degree of genealogical separation between nodes (i.e., how far back they shared a common ancestor). To test this, we introduced functions like: $0.05s$ ,where s denotes the separation index. While this resulted in an initial, favorable branching effect, the unbounded growth of the radius eventually forced nodes to the periphery, leading to a random and highly disorganized spatial distribution rather than structured growth. A final attempt involved combining both factors into more intricate functions, such as:  $1-e^{-0.2t-0.1s}$ and $(\tanh(0.2t-0.5s)+1)/2$. The models generated by these complex functions were convoluted and were hard to interpret, confirming that this level of mathematical complexity was ultimately counterproductive to achieving the desired formation.
The most effective solution was determined to be the simplest. We thus adopted a constant repulsion radius of $1/10$. This approach successfully achieved the primary goal, effectively preventing node overlap while simultaneously fostering the emergence of clear, distinct, and well-separated branching structures.

\begin{figure}
\caption{ Images of cloud model ran with various different equations for the radius of cloud of repulsion function $f(t,s)$, along with a density heatmap with $t$ for time and $s$ for degree of cousin separation.}
\begin{subfigure}{0.44\linewidth}
\includegraphics[width=0.9\linewidth]{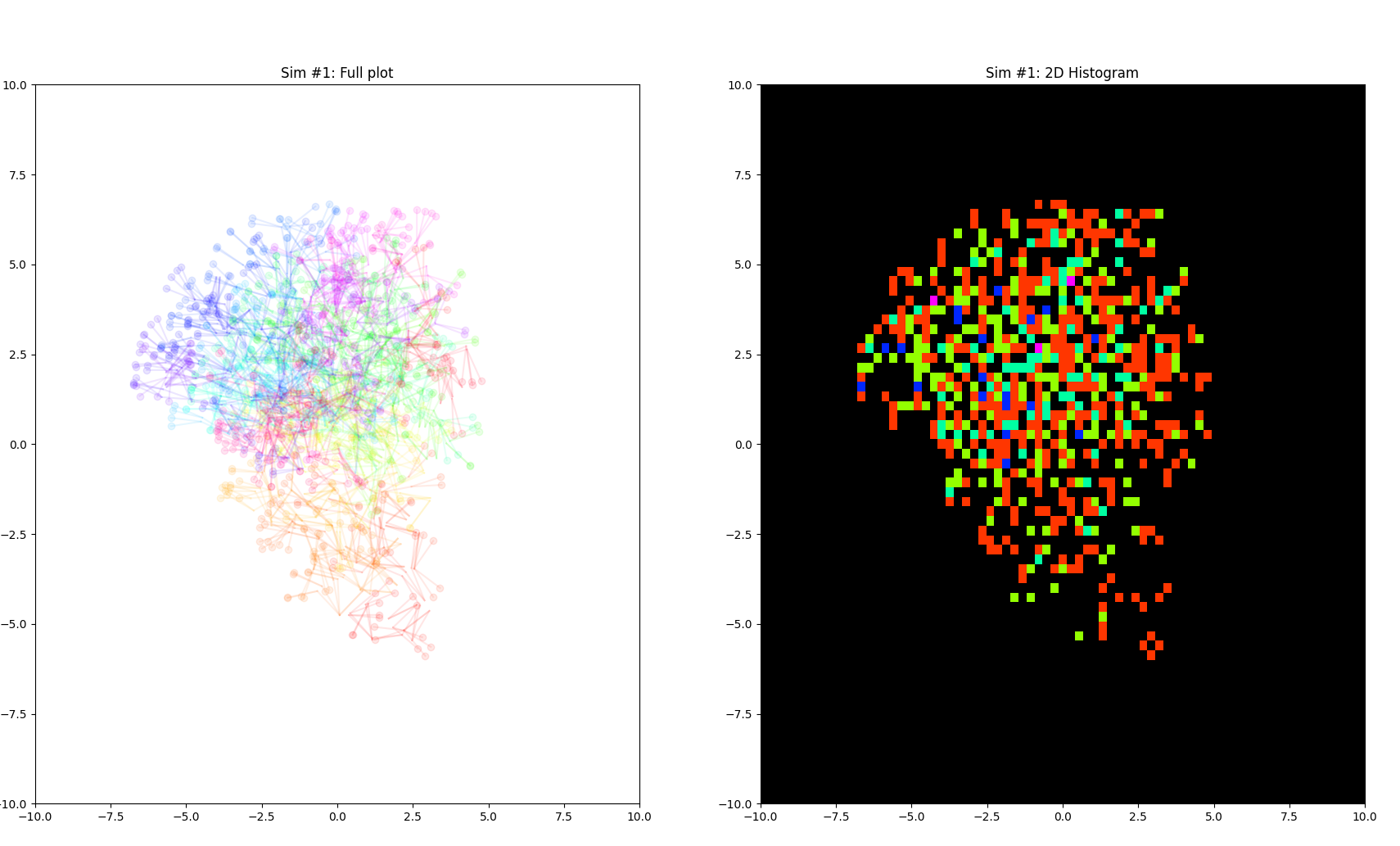} 
\caption{Radius of cloud $f(t,s) = (9/10)^t$,}
\label{fig:subim1}
\end{subfigure}
\begin{subfigure}{0.44\linewidth}
\includegraphics[width=0.9\linewidth]{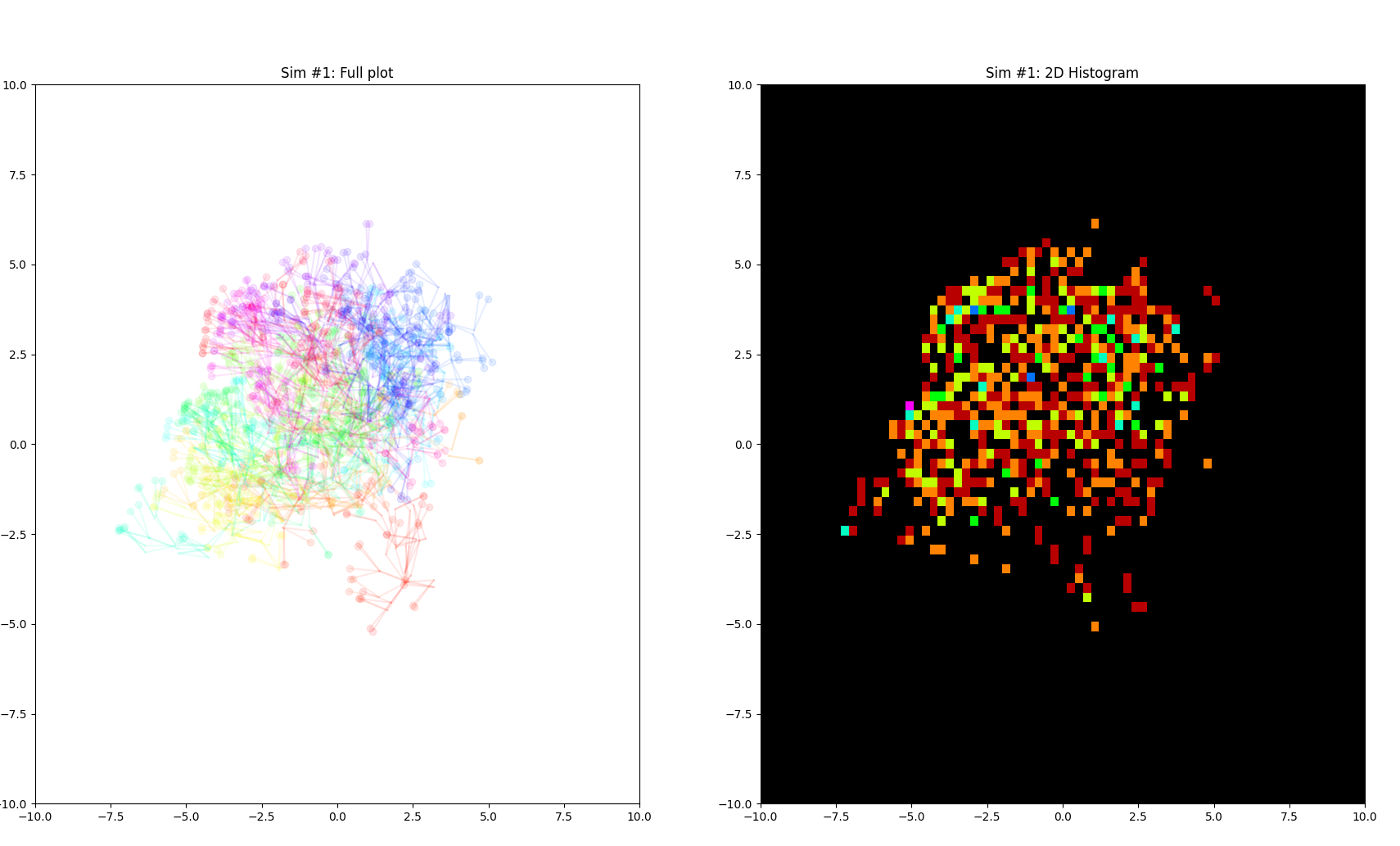}
\caption{Radius of cloud $f(t,s) = 0.05s$}
\label{fig:subim2}
\end{subfigure}
\begin{subfigure}{0.44\linewidth}
\includegraphics[width=0.9\linewidth]{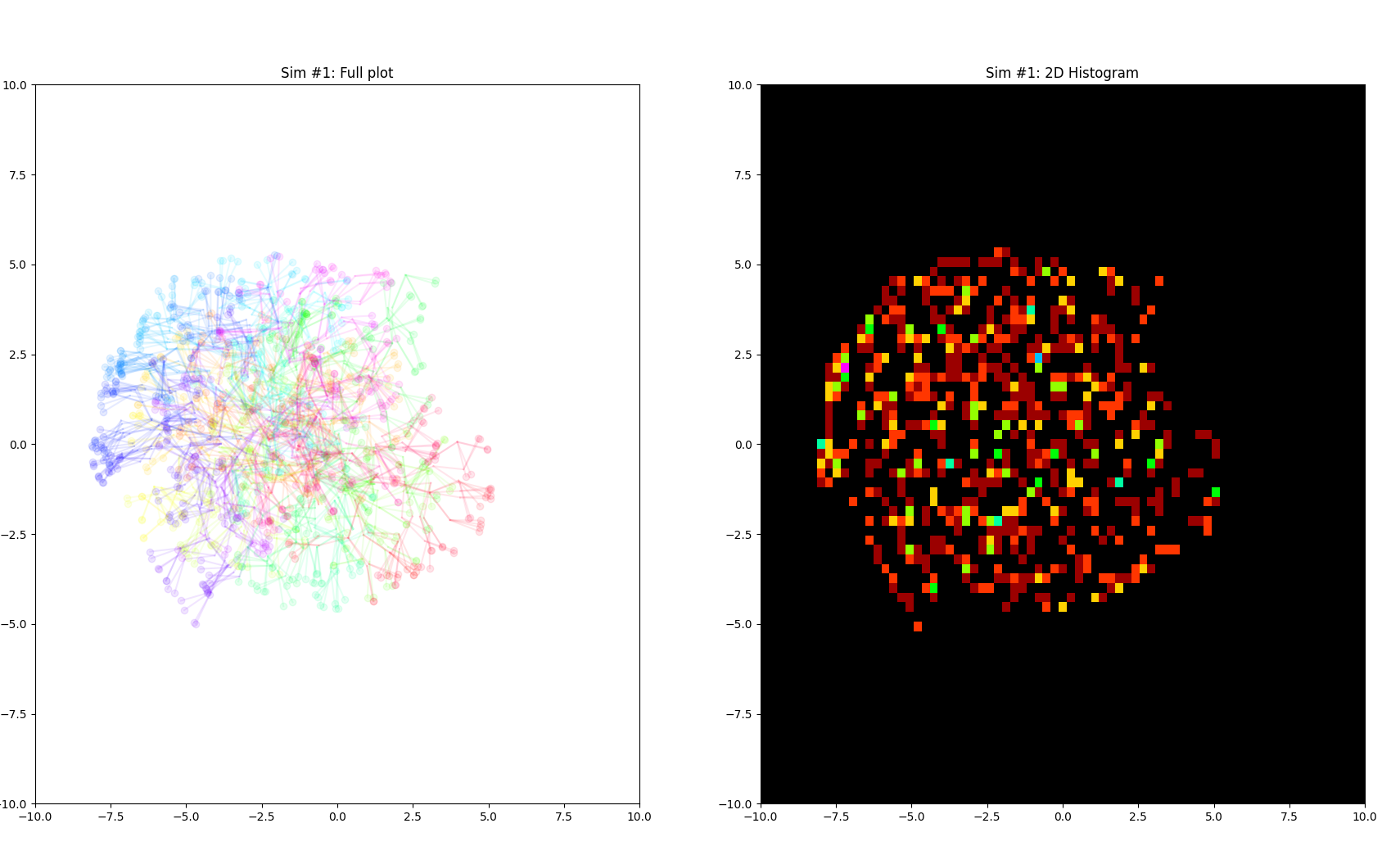}
\caption{Radius of cloud $f(t,s) = 1-e^{-0.2t-0.1s}$}
\label{fig:subim3}
\end{subfigure}
\begin{subfigure}{0.44\linewidth}
\includegraphics[width=0.9\linewidth]{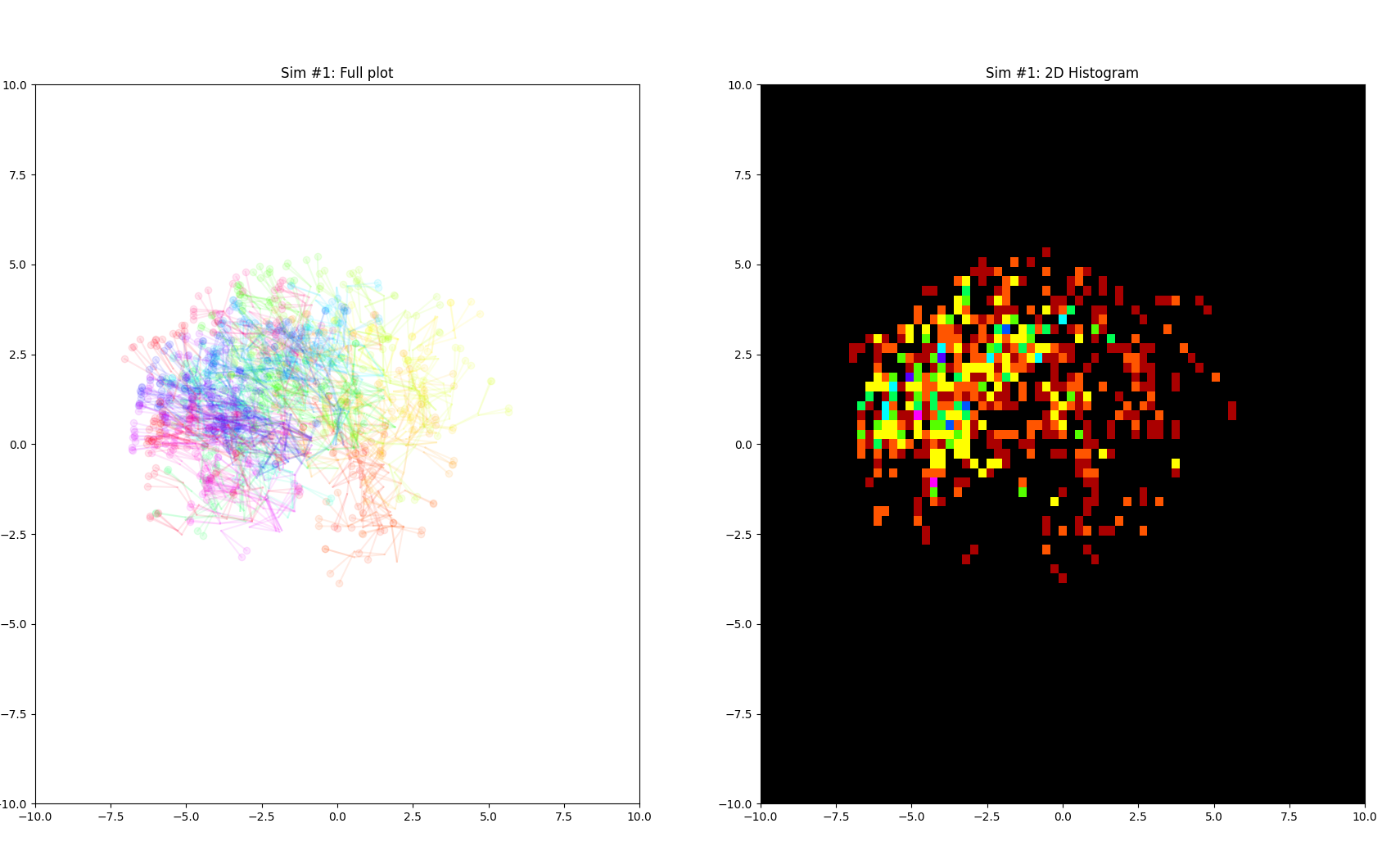}
\caption{Radius of cloud $f(t,s) = 1/10$}
\label{fig:subim4}
\end{subfigure}
\label{fig:image1}
\end{figure}

\section{Findings}\label{findings}
The models we proposed in the previous sections are stochastic processes. Each trial of one of these models results in a slightly different pattern (arrangement) of nodes at any given time. In order to discuss the findings from the models, we choose to perform nine trials of each particular model and observe the results at a fixed time (generation) point at the end. To observe the results, for each trial, we plot two images of the pattern of our simulation.  Of each pair, the left image, with the white background, includes the final position and path of each node alive when the simulation was terminated. The right image, with the black background, is the accompanying 2D density plots that show the spatial concentration of nodes (blue being the highest concentration). This density plot reveals the final pattern of the bacterial locations within the area visualized without their path history. 

In the equal weight model, all constituent models had the same weight in the combined model. However, when favoring a specific individual model, that model's weight was doubled (assigned a weight of two), with all other models receiving a weight of one in the combined model. Additionally, every simulation was conducted with consistent parameters: an angle variation of fifteen degrees, two children for each node, a total duration of ten generations, and the initial three generations generated completely randomly.

\subsection*{Equal Weight}

\begin{figure}
    \centering
    \includegraphics[width=0.9\linewidth]{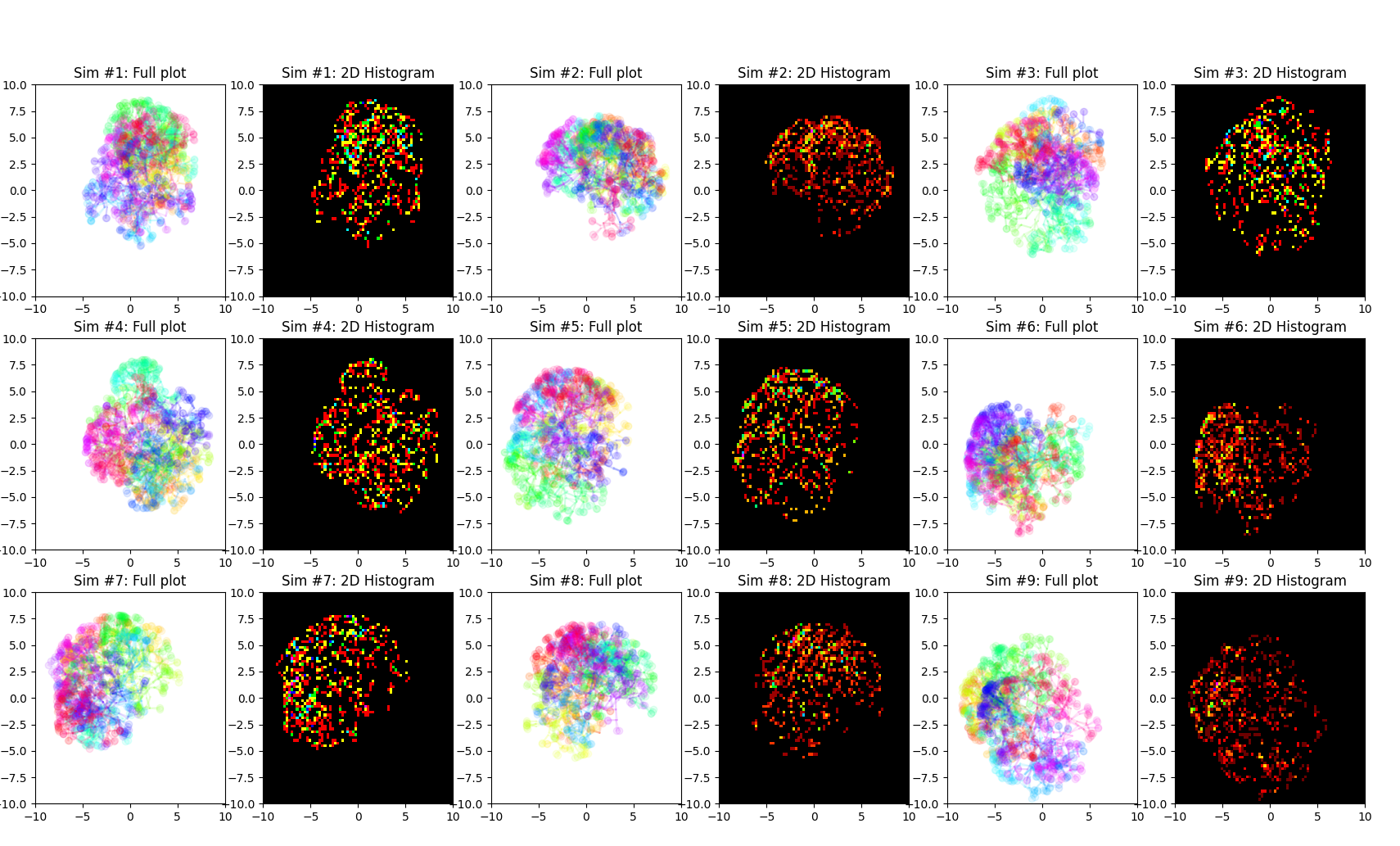}
    \label{fig:enter-label}
    \captionof{figure}{Simulations for nine trials of the combined model using the Equal Weight Model.}
\end{figure}

When each model is given equal weight, the nodes branch out like fireworks, spreading widely. Nodes from the same parent, however, tend to move in the same general direction. This model rarely folds in on itself, with nodes almost always heading outward. The resulting shape is a limiting form that resembles a circle, similar to a random generation. However, this model stands out due to its distinct bulges, which add a unique level of abstraction to the graph.

\subsection*{Favoring Density Model}

\begin{figure}
    \centering
    \includegraphics[width=0.9\linewidth]{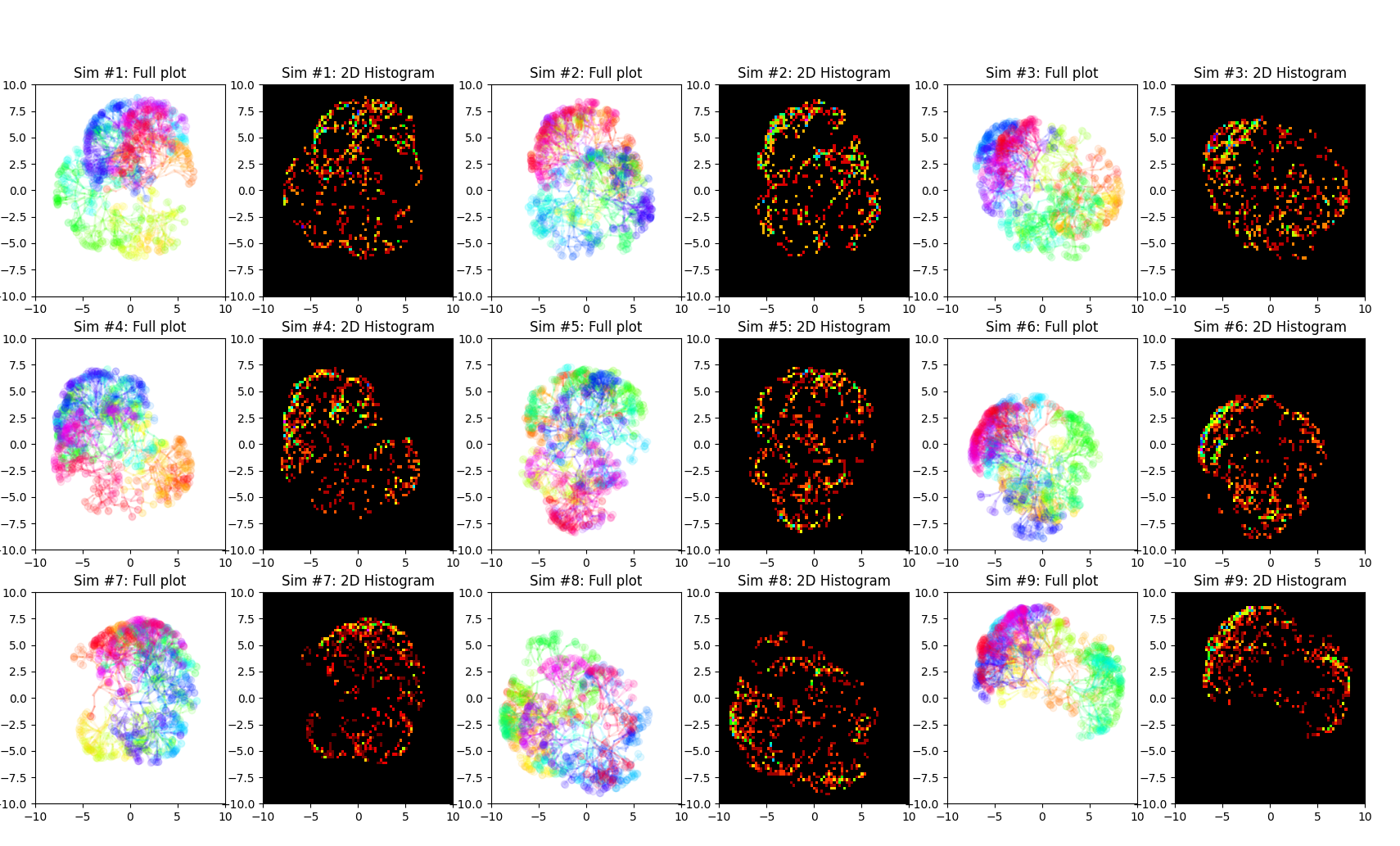}
    \label{fig:enter-label}
    \caption{Simulations for nine trials of the combined model using the Favoring Density Model.}
\end{figure}
When the models are weighted to favor density model, the resulting structures are characterized by clear, high-density protrusions extending from a central mass. This tree-like structure is a hallmark of many real-world bacterial growth patterns. Effective genealogical coherence is observed since child nodes move in the same general direction as their parents. This demonstrates the successful implementation of a genealogy-dependent factor in the BRW. Lineage-related particles (coded by color in the plots) maintain spatial proximity, forming the core of these visible branches. This links the growth process directly to the underlying family structure, which was the primary motivation for this research. The system achieves a balanced state of growth, as it simultaneously demonstrates internal genealogical coherence and undergoes external expansion, where nodes spread out and move toward larger groups, effectively contributing to the expansion of the colony boundary. This inter-group repulsion is a critical density-dependent feature. It suggests that the displacement distributions are not just affected by local density, but that particles from one lineage apply a repulsive force or create a high-density "barrier" that biases the random walk of particles from other lineages away from that area. This repulsion is what keeps the branches distinct rather than having them merge chaotically. The cone-like shape indicates that as a lineage grows, the core (the "trunk") maintains a high density, but the newer generations (the "rim" of the cone) are slightly more dispersed as they explore new space. This dispersion at the tip is necessary for continued growth.

\subsection*{Favoring Cloud Model}

\begin{figure}
    \centering
    \includegraphics[width=0.9\linewidth]{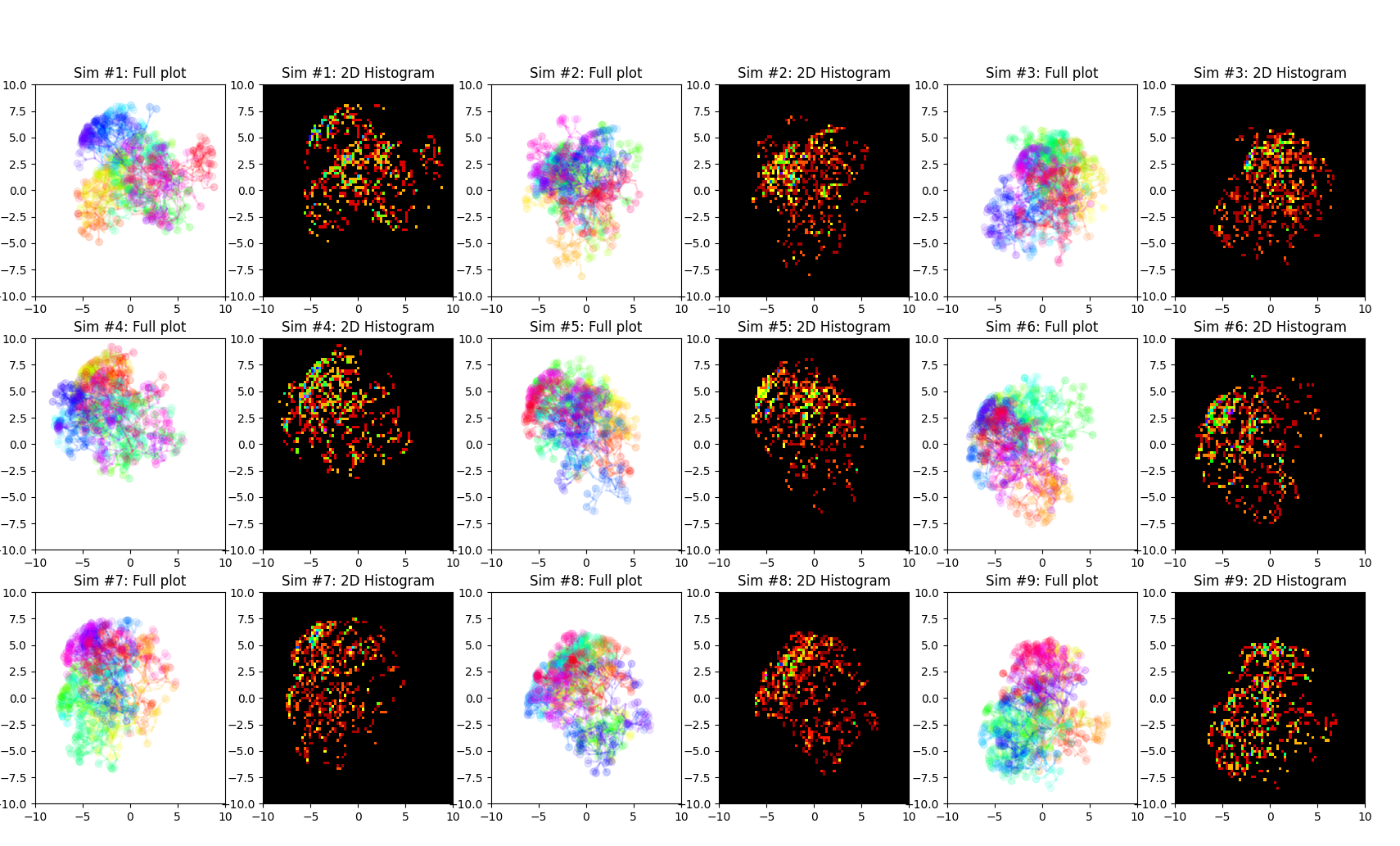}
    \label{fig:enter-label}
    \caption{Simulations for nine trials of the combined model using the Favoring Cloud Model. }
\end{figure}

The particles are scattered throughout the 2D space, giving the visual impression of clouds. This indicates that the displacement component of the repulsion from the cloud model is prominent. The overall shape of the cloud remains roughly circular and centered around the origin. The crucial observation is that nodes that are closely related do not stick together, and in fact, are pulled apart. The colors are heavily mixed and dispersed across the entire circular region confirming the fact that nodes with close genealogical connections are pulled apart. For example, a bright green lineage isn't confined to one edge or clump; its particles are scattered among the purples, reds, and blues. This visually confirms the lack of genealogical coherence: the particles' positions are primarily determined by the stochastic random walk rather than their lineage connection.

\subsection*{Favoring Cluster Model}
\begin{figure}
    \centering
    \includegraphics[width=0.9\linewidth]{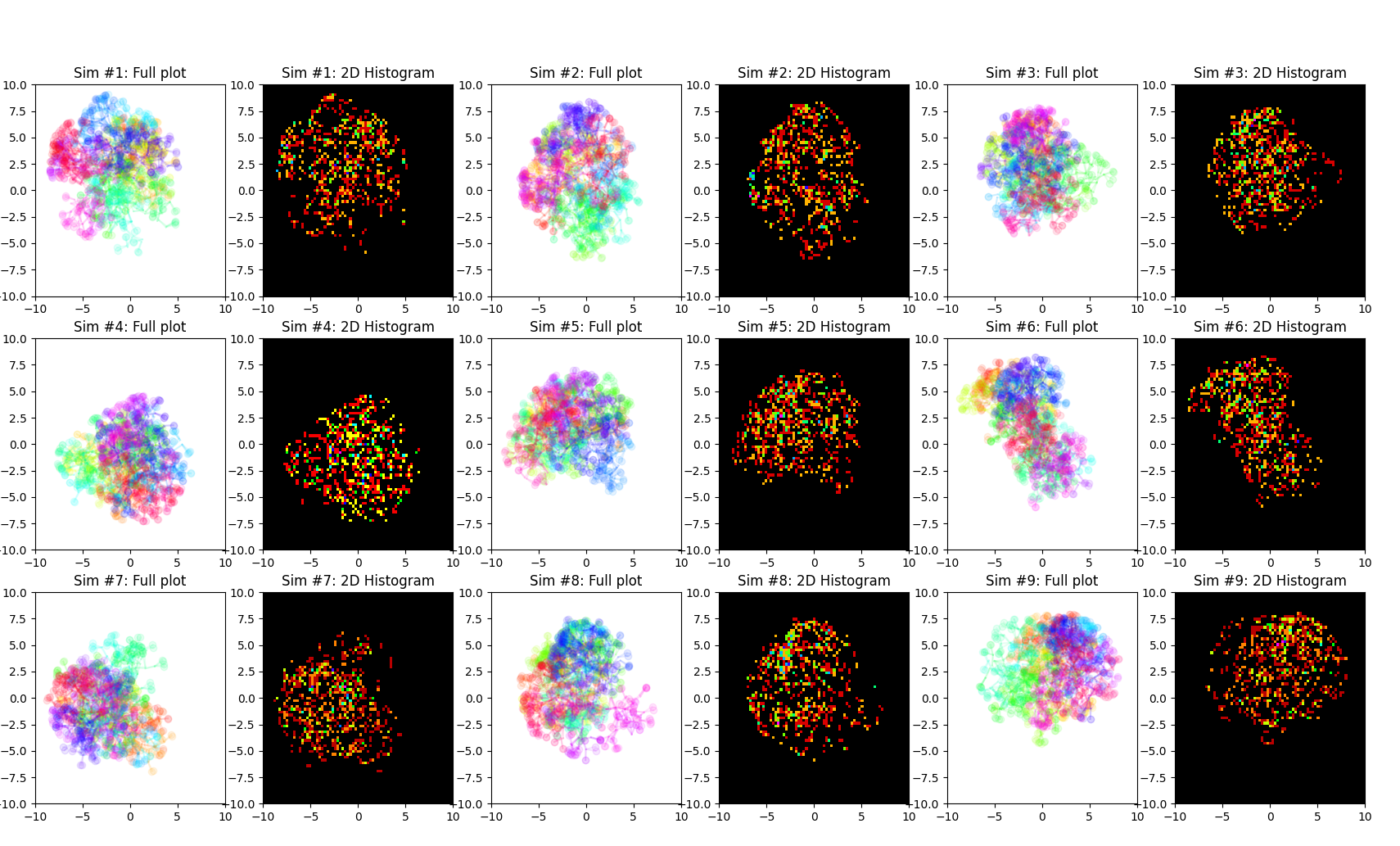}
    \label{fig:enter-label}
    \caption{Simulations for nine trials of the combined model using the Favoring Cluster Model.}
\end{figure}

The models successfully draw multiple nodes together, forming distinct clusters. This is a desired feature for simulating phenomena like bacterial colonies, where localized growth leads to dense patches. The loose "grip" on the nodes allows for the necessary branching effect, meaning new particles are successfully created within the existing clusters, increasing local population density and contributing to the growing mass. The significant problem identified is the loss of a clear growth direction once the nodes become dense. When nodes "bunch up", the local density and crowding (which can be a proxy for resource depletion or contact inhibition in a physical system) do not effectively bias the outward movement to model bacterial growth patterns. This suggests that the random walk component's displacement distribution is not being sufficiently modulated by spatial factors.

\section{Conclusions}\label{conlcusions}

This project aimed to simulate bacterial growth using a code based on branching random walks. While we developed a robust, adaptable model capable of combining numerous influencing factors, ultimately, none of the movement biasing rules we explored successfully replicated the precise natural bacterial formations we targeted. The next step would be to validate these models by moving from theoretical exploration to data-driven parameter fitting. This would require acquiring and quantifying real-world bacterial growth patterns to establish target observables, such as fractal dimension or aggregate shape/branching statistics. We could then implement an algorithm to search the parameter space from each model and the weight parameters in the combined model efficiently. The resulting best-fit parameter set will identify optimal parameters and the weights in the combined models, thereby revealing the most appropriate way to incorporate genealogical inhomogeneities, along with the spatiotemporal inhomogeneities in the BRWs via our proposed models to have the closest fit to the natural bacterial formations.

Our research was significantly constrained by the technology used, specifically a 2023 MacBook Air with 8 GB of memory. Due to the exponential growth of the model's population, the runtime increased dramatically with each generation. For the equally weighted model, the program's slowdown is evident:

\begin{tabular}{|c|c|}
\hline
   Generations	& Runtime (Approx.)   \\
   \hline
   10	&1 second\\
12	& 5 seconds\\
14	& 81 seconds (1.35 minutes)\\
15	& 325 seconds (5.5 minutes)\\
\hline
\end{tabular}

Furthermore, the model's reliance on past genealogical information to calculate the next generation's location presented a major hurdle for implementing parallelization algorithms to improve performance.

A way to mitigate this limitation would be to introduce culling into the model, which would decrease the population growth and allow the models to run for more generations, and facilitate better analysis of the limiting behavior and how it changes when influencing factors are implemented. There are many ways to implement culling to still preserve the properties of interest of our model. For example, if we are interested in the approximate shape of the growth of the model, a random culling of each generation could be implemented, where a fixed proportion $\theta \in (0,1)$ of particles alive at generation $t$ is uniformly selected and killed. If we were interested in studying the location of the frontier of growth, we could allow particles that are the leaders (fast particles) to survive and cull the rest.

\section{Further research}\label{further research}
There are various further lines of mathematical research we want to explore, building on numerous genealogy-dependent and density-dependent discrete branching random walk models we have proposed and simulated here. First, we aim to use our computer simulations and the connections between discrete-time random walks and continuous-time diffusion equations to study the long-time behavior of populations undergoing inhomogeneous branching diffusion processes by simulating the appropriate scaled discrete-time inhomogeneous branching random walks for long times. One of the first gaps we uncovered in the mathematical literature that exists is the analysis of the front propagation of branching diffusion processes is where the diffusion (or branching) mechanism is genealogy-dependent. Our work would provide the first step in understanding the fronts for such processes since the mathematical results are not available.

\section*{Acknowledgements}
This research was conducted during the summer of 2025 as part of a ten-week Mentored Advanced Project (MAP) at Grinnell College. The four undergraduate students L. Ajax, B. Durham, C. Johnston, J. Zhang, undertook this research and contributed equally while working under the faculty mentorship of Prof. P. Hebbar.  

\section*{Data availability} The code used to generate the results in this study is available at the following web address: 
https://github.com/pratima-hebbar/BranchingRandomWalks-MAP2025

\bibliographystyle{ieeetr} 
\bibliography{refrences}

\end{document}